\documentstyle[prl,aps,twocolumn,epsf]{revtex}

\begin{document}
\draft

\twocolumn[\hsize\textwidth%
\columnwidth\hsize\csname@twocolumnfalse\endcsname

\title{\bf Mound formation in nonequilibrium surface growth morphology 
does not necessarily imply a Schwoebel instability}

\author{S. Das Sarma, P. Punyindu, and Z. Toroczkai}
\address{Department of Physics, University of Maryland, College Park,
MD 20742-4111}

\date{August 1, 1999}
\maketitle

\begin{abstract}
We demonstrate, using well-established 
nonequilibrium growth models, that mound formation in
the dynamical surface growth morphology does not necessarily 
imply a surface edge diffusion bias
(``the Schwoebel barrier'') as has been almost universally 
accepted in the literature.
We find mounded morphologies in several 
nonequilibrium growth models which incorporate no Schwoebel barrier.
Our work should lead to a critical re-evaluation 
of recent experimental observations of mounded 
morphologies which have been theoretically interpreted in terms of
Schwoebel barrier effects.
\end{abstract}

\vskip 1pc]
\narrowtext
In vacuum deposition growth of thin films or epitaxial layers 
(e.g. MBE) it is common \cite{1} to find mound
formation in the evolving dynamical surface growth morphology.
Although the details of the mounded morphology could differ
considerably depending on the systems and growth conditions,
the basic mounding phenomenon in surface growth has been
reported in a large number of recent experimental publications
\cite{1}.
The typical experiment \cite{1} monitors vacuum deposition
growth on substrates using STM and/or AFM spectroscopies.
Growth mounds are observed
under typical MBE-type growth conditions, and the resultant 
mounded morphology is statistically analyzed 
by studying the dynamical surface
height $h({\bf r},t)$ as a function of the position
${\bf r}$ on the surface and growth time $t$.
Much attention has focused on this ubiquitous phenomenon
of mounding and the associated pattern formation during 
nonequilibrium surface growth for reasons of possible
technological interest (e.g. the possibility of producing 
controlled nanoscale thin film or interface patterns)
and fundamental interest (e.g. understanding nonequilibrium 
growth and pattern formation).

The theoretical interpretation of the mounding phenomenon has been
almost exclusively based
on the step-edge diffusion bias \cite{2} or the so-called
Schwoebel barrier \cite{3} effect
(also known as the Ehrlich-Schwoebel \cite{3},
ES, barrier).
The basic idea of the ES barrier-induced mounding 
(often referred to as an instability) 
is simple :
The ES effect produces an additional energy barrier for
diffusing adatoms on terraces from coming ``down'' toward
the substrate, thus probablistically inhibiting attachment
of atoms to lower or down-steps and enhancing their attachment
to upper or up-steps; the result is therefore mound formation
because deposited atoms cannot come down from upper to lower
terraces and so three-dimensional mounds or pyramids result
as atoms are deposited on the top of already existing terraces.

The physical picture underlying mounded growth under an ES
barrier is manifestly obvious, and clearly the existence of
an ES barrier is a {\it sufficient condition} \cite{2} for
mound formation in nonequilibrium surface growth. Our interest
in this Letter is to discuss the {\it necessary} condition for
mound formation in nonequilibrium surface growth morphology
--- more precisely, we want to ask the inverse question, namely,
whether the observation of mound formation  
requires the
existence of an ES barrier as
has been almost exclusively (and uncritically in our opinion)
accepted in the recent literature. Through concrete examples we
demonstrate rather compellingly that the mound formation in 
nonequilibrium surface growth morphology does {\it not}
necessarily imply the existence of an ES barrier, and we
contend (and results presented in this Letter establish)
that the recent experimental observations of mound formation in
nonequilibrium surface growth morphology should not be taken as
definitive evidence in favor of an ES barrier-induced
universal mechanism for pattern formation in surface growth.
Mound formation in nonequilibrium surface growth is 
a non-universal phenomenon, and could have very different 
underlying causes in different systems and situations.

Before presenting our results we point out that the possible 
nonuniversality in surface growth mound formation 
(i.e. mounds do not necessarily imply an ES barrier)
has recently been mentioned in at least two experimental
publications \cite{4,5} where it was emphasized that the 
mounded patterns seen on Si \cite{4} and
GaAs \cite{5}, InP \cite{5} surfaces during MBE growth
were not consistent with the phenomenology of a Schwoebel
instability. These papers \cite{4,5} have, however, been
essentially ignored in the literature, and the ES 
barrier-Schwoebel instability paradigm is by now so well-entrenched
in the literature that the experimental observations of 
mound formation during nonequilibrium growth are often forced 
to conform to the ES barrier scenario even when the 
resultant data analyses \cite{6} lead to the inconsistent 
conclusion about the non-existence of any ES barrier in
the systems under study \cite{6}.
There have been only two proposed mechanisms \cite{7,8}
in the literature which lead to mounding without any explicit
ES barrier: One of them invokes \cite{7} a preferential
attachment to up-steps compared with down-steps
(the so-called ``step-adatom'' attraction),
which, in effect, is equivalent to having an ES barrier
because the attachment probability to down-steps is
lower than that to up-steps exactly as it is in the
regular ES barrier case \cite{2,3} --- we therefore do not
distinguish it from the ES barrier mechanism, and in fact,
within the growth models we study, these
two mechanisms 
are physically and mathematically indistinguishable.
The second mounding alternative \cite{8}
is the so-called edge diffusion induced mounding,
where diffusion of adatoms
around cluster edges is shown to lead to mound
formation during nonequilibrium surface growth even in the
absence of any finite ES barrier. One of the concrete examples
we discuss below, the spectacular pyramidal pattern formation
(Fig. 3(c)) in the 2+1 dimensional (d) noise reduced
Wolf-Villain (WV) model \cite{9}, arises from such a
nonequilibrium edge diffusion effect (perhaps in a somewhat
unexpected context).
We also demonstrate, using the WV model and the
Das Sarma-Tamborenea (DT) model \cite{10}, that mound
formation during nonequilibrium surface growth is, in fact,
almost a generic feature of {\it limited mobility} growth
models \cite{9,10,11}, which typically have comparatively
large values of the roughness exponent \cite{11} ($\alpha$)
characterizing the growth morphology.
Below we demonstrate that mound formation
in surface morphology arising from this generic
``large $\alpha$'' effect (without any explicit ES barrier)
is qualitatively virtually indistinguishable from that in 
growth under an ES barrier.
Mound formation in the presence of strong edge diffusion
\cite{8} (as in the d=2+1 WV model in Fig. 3)
is, on the other hand, morphologically quite distinct from the
ES barrier- or the large $\alpha$- induced mound formation.

Our results are based on the extensively studied \cite{11} limited
mobility nonequilibrium WV \cite{9} and DT \cite{10} growth models.
Both models have been widely studied \cite{11}
in the context of kinetic surface roughening in nonequilibrium
solid-on-solid (SOS) epitaxial growth --- the interest in and the
importance of these models lie in the fact that these were the
first concrete physically motivated growth models falling
outside the well-known Edwards-Wilkinson-Kardar-Parisi-Zhang \cite{11}
generic universality class in kinetic surface roughening.
Both models involve random deposition of atoms on a square lattice
singular substrate (with a growth rate of 1 layer/sec.
where the growth rate defines the unit of time) under the
SOS constraint with no evaporation or desorption.
An incident atom can diffuse instantaneously before incorporation
if it satisfies certain diffusion rules which differ slightly
in the two models.
In the WV model the incident atom can diffuse within a
diffusion length $l$ (which is taken to be one with the
lattice constant being chosen as the length unit, i.e. only
nearest-neighbor diffusion, in all the results shown in this
paper --- larger values of $l$ do not change our conclusions)
in order to maximize its local coordination number or 
equivalently the number of nearest neighbor bonds it forms
with other atoms (if there are several possible final sites 
satisfying the maximum coordination condition equivalently
then the incident atom chooses one of those sites with equal
random probability and if no other site increases the local
coordination compared with the incident site then the 
atom stays at the incident site).
The DT model is similar to the WV model except for two 
crucial differences:
(1) only incident atoms with no lateral 
bonds (i.e. with the local coordination number of one ---
a nearest-neighbor bond to the atom below is necessary
to satisfy the SOS constraint) are allowed
to diffuse (all other deposited atoms, with one or more 
lateral bonds, are incorporated into the
growing film at their incident sites);
(2) the incident atoms move only to {\it increase} their 
local coordination number (and {\it not} to maximize it
as in the WV model) --- all possible incorporation sites
with finite lateral local coordination numbers are accepted
with random equal probability.
Although these two differences between the DT and the WV model
have turned out to be crucial in distinguishing their
{\it asymptotic} universality class, the two models
exhibit very similar growth behavior for a long transient
pre-asymptotic regime.
It is easy to incorporate \cite{12} an ES barrier in the
DT (or WV) model by introducing differential probabilities
$P_u$ and $P_l$ for adatom attachment to an upper and a lower 
step respectively --- the original DT model \cite{10} has
$P_u=P_l$, and an ES barrier is explicitly 
incorporated \cite{12} in the model by having 
$P_l < P_u \leq 1$. We call this situation 
the DT-ES model (we use $P_u = 1$ throughout with no
loss of generality). We also note, as mentioned above,
that within the DT-ES model the ES barrier 
\cite{2,3} ($P_l < P_u$) and the step-adatom attraction
\cite{7} ($P_u > P_l$) are manifestly equivalent, 
and we therefore do not consider them as separate mechanisms.
We note also that in some of our simulations below we 
have used the noise reduction technique \cite{12,13} 
which have earlier been successful in limited mobility growth
models in reducing the strong stochastic noise effect
through an effective coarse-graining procedure.

In Fig. 1 and 2 we present our d=1+1 
growth simulations, which demonstrate the point
we want to make in this Letter.  
We show in Fig. 1 the simulated growth morphologies at three
different times for four different situations, two of which
(Fig. 1(a),(b)) have finite ES barriers and the other
two (Fig. 1(c),(d)) do not. 
The important point we wish to emphasize is that, while the
four morphologies and their dynamical evolutions shown in 
Fig. 1 are quite distinct in their details, they all share one
crucial common feature: they all indicate mound formation
although the details of the mounds and the controlling length
scales are obviously quite different in the different cases.
Just the mere observation of mounded morphology, which is 
clearly present in Figs. 1(c),(d), thus does not necessarily 
imply the existence of an ES barrier. 
To further quantify the
mounding apparent in the simulated morphologies of Fig. 1
we show in Fig. 2 the calculated height-height correlation
function, 
$H(r) \sim \langle h({\bf x}) h({\bf r} + {\bf x})
           \rangle_{\bf x} ^{1/2}$,
along the surface for two different times.
\begin{figure}[htbp]
\hspace*{-1cm}
\epsfxsize=3.6 in
\epsfbox{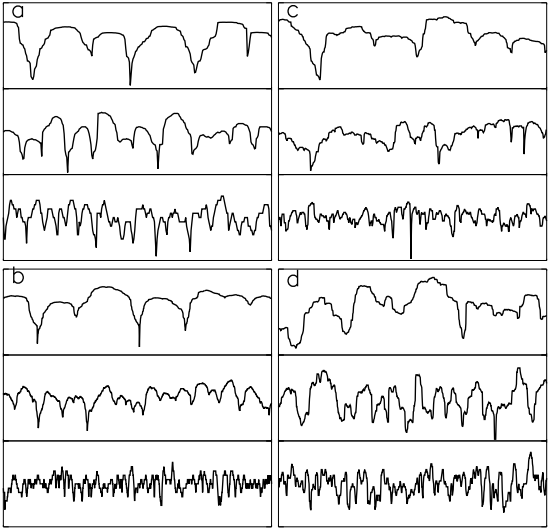}
\vspace*{-3.5cm}
\caption{
Dynamical morphologies at $10^2$, $10^4$ and $10^6$ monolayers (ML)
for (a) DT-ES with $P_l=0.5$, $P_u=1$;
(b) DT-ES with $P_l=0.9$, $P_u=1$;
(c) DT; and
(d) WV models.
}
\end{figure}
\noindent
All the calculated $H(r)$ show clear oscillations as a function
of $r$, which by definition implies mound formation.
It is indeed true that the presence of considerable stochastic
noise associated with the deposition process in the DT, WV models
make the $H(r)$-oscillations quite noisy, but there is no
questioning the fact that oscillations are present in $H(r)$
{\it even when there is no explicit ES barrier present in 
the growth model} (Figs. 2(c),(d)).
We have explicitly verified that such growth mounds (or
equivalently $H(r)$ oscillations) are absent in the growth 
models \cite{11} which correspond to the generic
Edwards-Wilkinson-Kardar-Parisi-Zhang universality class,
and arise only in the DT, WV limited mobility growth models
which exhibit non-generic behavior with a large value of
the roughness exponent $\alpha$.
In fact, the effective $\alpha$ in the DT, WV models
is essentially \cite{9,10,11} unity,
which is the same as what one expects in a naive theoretical
description of growth under the ES barrier
(although the underlying growth mechanisms are completely
different in the two situations).
We believe that any surface growth involving a 
``large'' roughness exponent
( $0.5 < \alpha \lesssim 1$)
will invariably show ``mounded'' morphology
independent of whether there is an ES barrier in the
system or not. We contend that this 
effectively large $\alpha$
is the physical origin for mounded morphology in 
semiconductor MBE growth where one expects the
surface diffusion driven
linear or nonlinear conserved fourth order
(in contrast to the generic second order) dynamical growth
universality \cite{11} class 
to apply which has the asymptotic exponent :
$\alpha$ (d=1+1) $\approx$ 1; 
$\alpha$ (d=2+1) $\approx$ 0.67 (nonlinear), 1 (linear).
One recent experimental paper \cite{5}, which reports
the observation of mounded GaAs and InP growth with
$\alpha \approx 0.5 - 0.6$, has explicitly made this case,
and all the reported mound formations \cite{1,6}
in semiconductor MBE growth are consistent with our 
contention of the mounds arising from
[as in our Fig. 1(c),(d)] a large effective roughness exponent
rather than a Schwoebel instability.
The calculated \cite{14} ES barrier on semiconductor 
surfaces are invariably small,
providing further support to our contention that 
mounding in semiconductor surface growth is not an 
ES barrier effect, but arises instead from the fourth order
growth equations \cite{9,10,11,12,13} which have large
roughness exponents. Two very recent experimental publications
\cite{R} have reached the same conclusion in non-semiconductor
MBE growth studies --- in these recent publications \cite{R}
spectacular mounded surface growth morphologies have been 
interpreted on the basis of the fourth order 
conserved growth equations \cite{9,10,11,12,13}. 
\begin{figure}[htbp]
\hspace*{-1cm}
\epsfxsize=3.6 in
\epsfbox{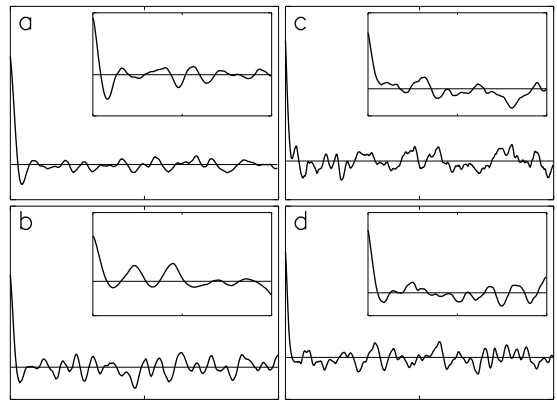}
\vspace*{-2.5cm}
\caption{
The height-height correlation function $H(r)$ at $10^2$ML
(main plots) and $10^4$ML (insets) corresponding respectively to the
morphologies in Fig. 1.
}
\end{figure}

Finally, in Fig. 3 and 4 we present our results for
the physically more relevant d=2+1 nonequilibrium surface growth.
In Fig. 3(a)-(c) we show the growth morphologies for
the DT-ES, DT,
and the noise-reduced WV model, 
respectively whereas in the main Fig. 3 we show the scaled
height-height correlation function. 
It is apparent that all three models 
(one with an ES barrier and the other two without)
have qualitatively similar oscillations in $H(r)$
indicating mounded growth, and the differences between
the growth models are purely quantitative.
Thus we come to the same conclusion: 
mound formation, by itself, does not imply the 
existence of an ES barrier; 
the details of the morphology 
obviously will depend on the existence (or not)
of an ES barrier. 
We note that the effective values
of the roughness exponent are very similar in 
Fig. 3(a) and (b), both being approximately
$\alpha \sim 0.5$
(far below the asymptotic value $\alpha \approx 1$
expected in the ES barrier growth --- we have verified that
this asymptotic $\alpha \approx 1$ is achieved in our 
simulations at an astronomically long time of $10^9$ layers).
\begin{figure}[htbp]
\hspace*{-1cm}
\epsfxsize=3.6 in
\epsfbox{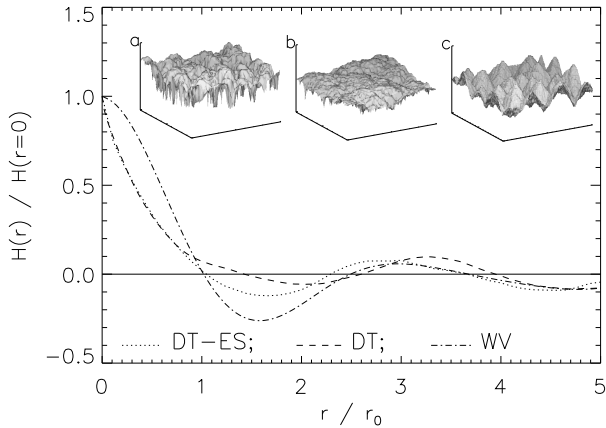}
\vspace*{-5cm}
\caption{
The scaled $H(r)$ correlation functions corresponding to the
morphologies shown in the insets. ($r_0 \equiv$ mound radius.)
Insets: Morphologies from the (a) DT-ES with $P_l=0.5$, $P_u=1$;
(b) DT; and (c) noise reduced WV models.
}
\end{figure}
\noindent
The most astonishing result we show in Fig. 3 is the
spectacular pyramidal mound formation in the
d=2+1 noise reduced WV model (without any
ES barrier). 
The strikingly regular pyramidal pattern
formation (Fig. 3(c)) in our noise reduced WV model
in fact has a magic slope and
strong coarsening behavior.
The pattern is very reminiscent of the theoretical
growth model studied earlier in ref. \cite{15} in the context of
nonequilibrium growth under an ES barrier where very
similar patterns with slope selection were proposed as
a generic scenario for growth under a Schwoebel instability.
In our case of the noise reduced d=2+1 WV model 
of Fig. 3(c), there is no ES barrier, but there is strong
cluster-edge diffusion as explained 
schematically in Fig. 4. 
This strong edge
diffusion (which obviously cannot happen in 1+1 dimensional growth)
arises in the WV model (but {\it not} in the DT model)
from the hopping of adatoms which have finite lateral 
nearest neighbor bonds (and are therefore the edge
atoms in a cluster).
This edge diffusion leads to an ``uphill'' surface current
(discussed in entirely different contexts in [8]),
which leads to the formation of the slope-selected 
pyramidal patterned growth morphology.
While noise reduction enhances the edge current
strengthening the pattern formation
(the uphill current is extremely weak in the 
ordinary WV model due to the strong suppression by
the deposition shot noise),
our results of Fig. 3 estabish compellingly
that the WV model in d=2+1 is, in fact,
unstable (uphill current) in contrast to the situation
in d=1+1.  
Thus, the WV model belongs to 
totally different universality classes in d=1+1 and 2+1
dimensions!
We mention that in (unphysical) higher (e.g. d=3+1, 4+1, etc.)
dimensions, the WV model would be even more unstable, 
forming even stronger mounds since the edge diffusion
effects will increase substantially in higher dimensions
due to the possibility of many more configurations of
nearest-neighbor bonding. 
We have therefore provided the explanation for the
long-standing puzzle of an instability in high-dimensional
(d $>$ 2+1) WV model simulations which were reported
\cite{16} in the literature some years ago.
More details on this phenomenon will be published elsewhere \cite{17}.
\begin{figure}[htbp]
\vspace*{-1cm}
\hspace*{-1cm}
\epsfxsize=3.3 in
\epsfbox{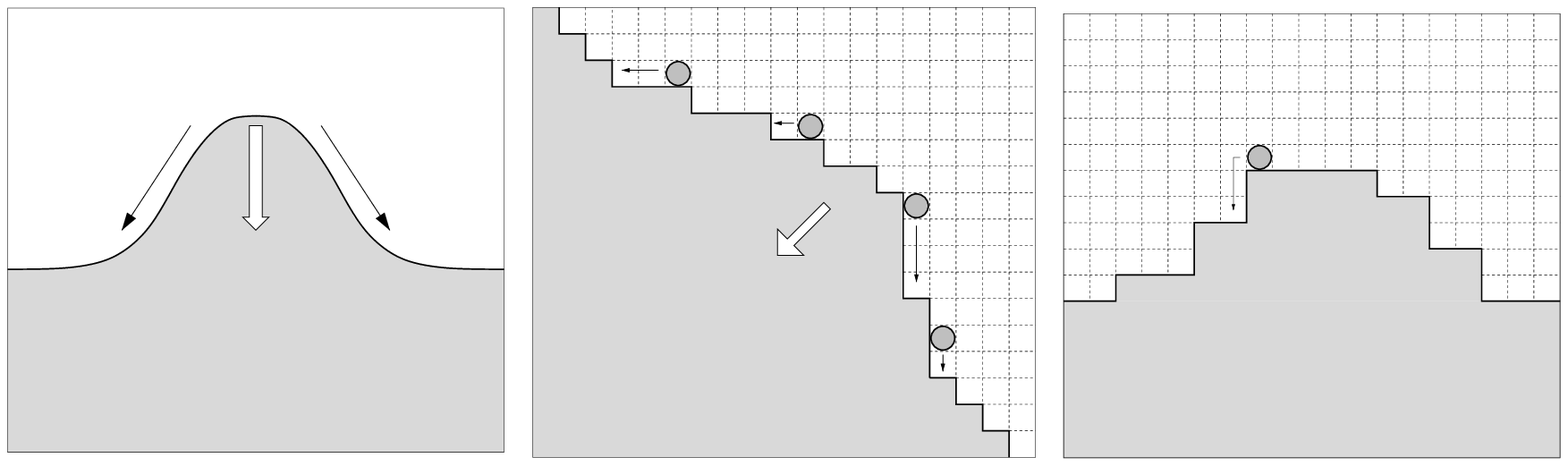}
\vspace*{-7cm}
\caption{
Schematic illustration of the instability caused by the
line tension in the step-edge. The view is from above, the shaded
region is the terrace of higher elevation.
Left panel shows a continuum picture while others represent the
same on a lattice for different step orientations.
}
\end{figure}

In conclusion, we have shown through concrete examples
that, while a
Schwoebel instability is certainly sufficient to cause 
mounded surface growth morphology, the reverse
(which has been almost universally assumed in the literature)
is simply not true :
an ES barrier is by no means necessary to produce mounds, 
and mound formation in nonequilibrium surface growth morphology 
does not necessarily imply the existence of a Schwoebel
instability. In particular, we show that a large roughness
exponent (without any ES barrier) as in the fourth order
conserved growth universality class \cite{9,10,11,12,13}
produces mounded growth morphologies which are
indistinguishable from the ES barrier effect.

\end{document}